\documentclass[narrowdisplay]{elsart5p}
\usepackage{color}
\usepackage{amssymb,amsmath}
\usepackage{graphicx}

\newcommand{\fr}{\frac}
\newcommand{\beq}{\begin{equation}}
\newcommand{\eeq}{\end{equation}}
\newcommand{\bea}{\begin{eqnarray}}
\newcommand{\eea}{\end{eqnarray}}
\journal{Physics Letters B}
\begin{document}
\begin{frontmatter}
\begin{flushright}
JLAB-THY-07-737\end{flushright}
\title{
   Analysis of Negative Parity Baryon  Photoproduction Amplitudes\\
in the $1/N_c$ Expansion}
\vspace{0.4cm}
\author[ato,conicet,favaloro]{N. N. Scoccola}  \ead{scoccola@tandar.cnea.gov.ar}$\negthinspace\negthinspace$,
\author[hampton,jlab]{J. L. Goity} \ead{goity@jlab.org}$\negthinspace\negthinspace$, and
\author[liege]{N. Matagne}  \ead{nmatagne@ulg.ac.be}
\vspace{0.4cm}
\address[ato]{Physics Depart., Comisi\'on Nacional de Energ\'{\i}a
At\'omica, (1429) Buenos Aires, Argentina}
\address[conicet]{CONICET, Rivadavia 1917, (1033) Buenos Aires, Argentina}
\address[favaloro]{Universidad Favaloro, Sol{\'\i}s 453, (1078) Buenos Aires,
Argentina}
\address[hampton]{Department of Physics, Hampton University, Hampton, VA 23668, USA}
\address[jlab]{Thomas Jefferson National Accelerator Facility, Newport News, VA 23606, USA}
\address[liege]{University of Li\`ege, Institute of Physics B5, Sart Tilman,
B-4000 Li\`ege 1, Belgium}
\date{\today}
\begin{abstract}
We study the photoproduction helicity amplitudes of
negative  parity  baryons in the context of the $1/N_c$
expansion of QCD. A complete analysis to  next-to-leading order is carried out.
The results show  sub-leading effects to be within the magnitude expected from the $1/N_c$
power counting.  They also show significant deviations from the
quark model,  in particular   the need for   2-body effects.
\end{abstract}\end{frontmatter}
\maketitle
\section{Introduction}

In this letter,  the photoproduction  helicity amplitudes of the
first excited negative parity baryons are analyzed in the
framework of the $1/N_c$ expansion of QCD \cite{tHo74}. Those
baryons   belong  to the $[{\bf 20'},1^-]$ multiplet of
$SU(4)\times O(3)$, where ${\bf 20'}$ is   the mixed symmetric
representation of $SU(4)$ (non-strange states in the $SU(6)$ ${\bf
70}$-plet). In terms of  masses and widths as well as
electromagnetic helicity amplitudes, this is the best known
multiplet of excited baryons.   In the  $1/N_c$ expansion,  the
masses were analyzed in Refs.  \cite{CCGL,GSS02}, and the strong
transition partial  widths were analyzed in Refs.
\cite{Carone,GSS1}.  The photoproduction helicity amplitudes have
been studied for more than  forty years in many works,
predominantly  using  constituent quark models
\cite{Capstick:2000qj},  the related single-quark transition model
based on $SU(6)_W$ symmetry \cite{Clo79}, and dispersion
approaches \cite{Dispersion}. In the $1/N_c$ expansion  the first
analysis of negative parity baryon helicity amplitudes  was
carried out by Carlson and Carone \cite{carlson}.  Positive parity baryon
helicity amplitudes have  also been also analyzed in recent work
  \cite{GSelm}.  Some model independent relations for helicity amplitudes have been obtained in
Ref. \cite{Cohen:2004bk}. The present work extends the analysis in Ref.
\cite{carlson}  by systematically  building a complete basis of
current operators to sub-leading order in the $1/N_c$  expansion,
 and by presenting and discussing the results in terms of the
multipole contributions to each helicity amplitude. We will
compare our analysis with that of Ref. \cite{carlson} in the
discussion of results.

The photoproduction helicity amplitudes are defined by the following matrix elements:
 \begin{equation}
A_\lambda = - \sqrt{\fr{2\pi\alpha}{\omega}}\;
\eta(B^*)\;
  \langle B^*, \lambda \mid
{\vec\epsilon}_{+1}\cdot\vec{J}(\omega \hat{z})
\mid N, \lambda-1 \rangle.
\label{uno}
\end{equation}
They  correspond to  the  standard definition as used by the Particle Data Group
 \cite{PDG06},  which includes a sign factor $\eta(B^*)$ that stems from the strong
decay amplitude    of the excited baryon to a $\pi N$ state.
The amplitudes   in Eqn. (1) are  independent of the phase
conventions used to define the excited states. The sign factors
are on the other hand convention dependent. Here $N$ and
$B^*$  denote respectively the initial nucleon and the final
excited baryon,   $\lambda=1/2$ or
$3/2$ is the helicity defined along  the $\hat{z}$-axis  which
coincides with the photon momentum,  ${\vec\epsilon}_{+1}$ is the
photon's polarization vector for helicity $+1$,  and  $\omega=(M_{B^*}^2-M_N^2)/2 M_{B^*}$ is the
photon energy in the rest frame of $B^*$.  In the $1/N_c$
expansion,  the electromagnetic current $\vec{J}$ is represented as
a linear combination of effective multipole current
operators with the most general form:
\beq \left(k^{[L']}
{\cal B}^{[LI]}\right)^{[1I]}, \label{dos} \eeq
where the upper
scripts display the angular momentum and isospin, and throughout
the neutral component, \emph{i.e.}  $I_3=0$, is taken. The $O(3)$ tensor $k^{[L']} $
is expressed in terms of  spherical harmonics of the photon momentum,  and  ${\cal B}^{[L
I]}=\left(\xi^{(\ell)} {\cal{G}}^{[\ell' I]}\right)^{[LI]}$ are
baryonic  operators.     $ \xi^{(\ell)}$ is  the
tensor associated with the transition from the $\ell=0$  $O(3)$ state of the nucleon to the  $O(3)$ state
of the excited baryon, and is normalized by its
reduced matrix element (RME) according to $\langle 0 \mid\mid \xi^{(\ell)}\mid\mid
\ell\rangle= \sqrt{2\ell+1}$ ($\ell=1$ in this work).  Finally,
${\cal{G}}^{[\ell' I]}$ is a spin-flavor tensor operator with
$I=0$ or $1$.   The parity selection rules imply  that the
helicity amplitudes for  photoproduction of the $[{\bf 20'}, 1^-]$  states can
only contain $E1$, $M2$ and $E3$ multipoles.  The quantum number
$L$ in Eqn. (\ref{dos}) determines the multipole: $EL$ for $L=1,3$
and $M2$ for $L=2$. For the $EL$ multipoles the photon orbital
angular momentum $L'$ is $L'=L\pm 1$ and for $ML$ multipoles
$L'=L$.
The multipoles
are in addition classified according to their isospin, into
isoscalars and isovectors. For general $N_c$ the isovector and
isoscalar components of the electric charge can be generalized in
different ways \cite{Lebed:2004fj}. Here we consider them as being
both ${\cal O}(N_c^0)$, corresponding to the  assumption that
quark charges are $N_c$ independent.

The multipole components of the helicity amplitudes are expressed
in terms of the matrix elements of the effective operators as follows:
\begin{eqnarray}
A_\lambda^{ML}& &\hspace{-0.05cm} = \sqrt{\frac{3 \alpha N_c }{4 \omega} }  (-1)^{L+1}\; \eta(B^*) \
\sum_{n,I} g^{[L,\, I]}_{n,L} (\omega)  \!\! \\
& & \hspace{-0.05cm} \times \left<J^*, \lambda; I^*, I_3;S^*\!\mid
\left({\cal B}_n\right)^{[L,\, I]}_{[1,0]}
\mid \! 1/2 , \lambda-1 ; 1/2,  I_{3} \right>\nonumber,
\label{aml2}\\
A_\lambda^{EL}& &\hspace{-0.05cm} = \sqrt{\frac{3 \alpha N_c }{4 \omega} } \;(-1)^{L}\; \eta(B^*)  \\ & & \hspace{-0.05cm}\times
\sum_{ n, I}\; \left[ \sqrt{\frac{L+1}{2L+1}}\ g^{[L,\, I]}_{n,L-1}  (\omega)
                        + \sqrt{\frac{L}{2L+1}}\ g^{[L,\, I]}_{n,L+1}  (\omega)
                        \right]
\nonumber \\ & &  \hspace{-0.05cm}\times \left<J^*,
\lambda; I^*, I_3;S^*\!\mid
\left({\cal B}_n\right)^{[L,\,
I]}_{[1,0]} \mid \! 1/2 , \lambda-1 ; 1/2,  I_{3} \right>\nonumber,
\label{ael2}
\end{eqnarray}
where $J^*$, $I^*$ and $S^*$ denote the spin, isospin and quark-spin of the excited baryon
and the sum over $n$  is over all operators with  given $[L,I]$ quantum numbers.
The factor $\sqrt{N_c}$ appears as usual for transition matrix elements between excited
and ground state baryons \cite{Goity:2004pw}.
 In the electric multipoles we have a combination
of the coefficients $g^{[L,\, I]}_{n,L-1}$ and $g^{[L,\,
I]}_{n,L+1}$, and because the operators appearing in these
multipoles do not appear in the magnetic multipoles, we may as
well replace that combination of coefficients by a single term
without any loss of generality. Thus, in what follows we will only
keep $g^{[L,\, I]}_{n,L-1}$. These  and the coefficients $g^{[L,\, I]}_{n,L} (\omega)$
are going to be determined by  fits to the empirical helicity amplitudes.

It is convenient to express these matrix elements in terms of
reduced matrix elements (RMEs)  via the Wigner-Eckart
theorem:
\begin{eqnarray}
\lefteqn{\left<J^*, \,\lambda;\, I^*,\, I_3;S^*\mid
\left({\cal B}_n\right)^{[L,\,I]}_{[1,0]}
\mid  1/2,\, \lambda-1 ; \,1 / 2 ,\, I_{3} \right>
 =} \nonumber \\ & &  \frac{(-1)^{L + I + J^{*} + I^{*} - 1}}{\sqrt{(  2 I^{*} + 1) ( 2 J^{*} +1)} }\;
\left < L ,\, 1  ; \,1/2 , \,\lambda-1 \mid J^{*} , \, \lambda \right> \nonumber \\ & & \times
 \left< I ,\, 0  ; 1/2 ,\, I_{3} \mid I^{*} , \,I_{3}\right >\;
   \left< J^{*};I^{*}; S^*  || {\cal B}_{n}^{[L,\, I]} || 1/2,1/2  \right>.
\label{unos}
\end{eqnarray}
 If one wishes, one can further express  the  RMEs of the baryonic operators
 in terms of  RMEs  involving only the spin-flavor pieces of  those operators \cite{GSelm}.
 
For the purpose of carrying out the group theoretical calculations,  and without any loss of
generality, one can consider that the $[{\bf 20'},1^-]$ baryon states are made of a ground state
core composed of $N_c-1$ quarks coupled to an excited quark.
The states can then be expressed as follows \cite{CCGL,Goity}:
\begin{eqnarray}
\lefteqn{| J,J_3 ; I,I_3 ; S \rangle    =\! \! 
  \sum_{m,s_3,i_3,\eta}\!\!\! \left< \ell,\,m;S,\,J_3-m\!\mid \! J, J_3\right>c_{\mathrm{MS}}(I,S,\eta)}\nonumber \\
&& \times\left< S_c,\,S_3-s_3;\,1/2,\,s_3\!\mid\! S,\,S_3\right>\;
\left< I_c,\,I_3-i_3;\,1/2,\,i_3\! \mid\! I,\,I_3\right> \nonumber \\
&& \times \mid\! S_c ,  S_3-s_3 ; I_c=S_c, I_3-i_3 \rangle\, | 1/2, s_3 ; 1/2, i_3   \rangle\, | 1 , m \rangle,
\end{eqnarray}
where   $\ell=1$,  $\eta=\pm 1/2$, $S_c=I_c=S+\eta$ are the spin and the isospin of the core,  and $c_{\mathrm{MS}}(I,S,\eta)$ are isoscalar
factors of the permutation group of $N_c$ particles  \cite{MS}, which for the mixed symmetric representation $[N_c-1,1]$
can be found in Ref. \cite{CCGL}.
In the following,  the generators of $SU(4)$ which act on the core will carry a
subscript $c$, while operators acting on the excited quark will be denoted in
lower case. For $N_c=3$,  the states contained in the $[{\bf 20'}, 1^-]$  are as follows:
two $N$ states with $J^*=1/2$, two with $J^*=3/2$ and one with $J^*=5/2$, and
one $\Delta$ with $J^*=1/2$ and one with $J^*=3/2$.
There are two mixing angles, $\theta_1$ for the  pair of excited $N$ states
with $J^*=1/2$,
and $\theta_3$ for the $N$ pair with  $J^*=3/2$.  The mixing  angles are defined in the standard fashion   \cite{CCGL},
and have been determined in different ways. In the  $1/N_c$  expansion in particular,  they can be obtained from an analysis
of the  masses \cite{CCGL},  and more precisely from  analyzing
strong transitions \cite{GSS1}. We use the latter in this work.

The basis of baryon   operators ${\cal B}$ can be built using  leading and sub-leading spin-flavor operators by following a procedure similar to that described
in Ref. \cite{GSS1} for the case of the strong decays.
 The basis used in
 this work is depicted in Table 1, which indicates the multipole to which the
 operator contributes and the order in $1/N_c$.
More specifically,  a baryonic  operator  ${\cal B}$  is given by the corresponding operator
 in the basis of Table 1 multiplied by  a scaling factor $\alpha$, depicted in the last column of
 the table, which is introduced in order for the operator to have matrix elements of natural size. This
 factor $\alpha$ is chosen in such a way that the largest RME of  the operator  ${\cal B}$
 is  equal to $1$ ($1/3$) if the operator is
 ${\cal O}(N_c^0)$ (${\cal O}(1/N_c)$). This allows one easily see the importance of the different
 operators by just looking at the magnitude of their coefficients.  At leading order (LO) in $1/N_c$
 there are a total of eight operators, one $E1$ and
 one $M2$ isoscalars, and three $E1$, two $M2$ and one $E3$ isovectors.
 It is important to emphasize that this distribution in the different multipoles
 is basis independent. At sub-leading order (NLO),  there are eleven new operators.
 This exhausts the basis because the number of helicity amplitudes for the
 photoproduction of the $[{\bf 20'}, 1^-]$ baryons is equal to nineteen. The analysis shows,
 therefore,  that neither sub-sub-leading operators nor three-body operators are needed
 for a full description of the helicity amplitudes.  This in particular means that there
 is no way of  sorting  out such contributions.

\begin{table}[b]
\caption{Baryon operator basis. The upper labels $^{[L,I]}$ denote angular momentum
and isospin and how these are coupled.  The  NLO operators  $E1^{(1)}_5$,   $E1^{(1)}_6$, and  $M2^{(1)}_4$
involve linear combinations with LO  operators in order to eliminate the LO  component.  }
\vspace*{2mm}
\begin{center}
\begin{tabular}{clcc}
\hline \hline
   \multicolumn{2}{c}{Operator}  &  Order & \hspace*{.2cm} Type \hspace*{.2cm} \\
\hline \hline
$E1^{(0)}_1=$ & $\left( \xi^{[1,0]} s\right)^{[1,0]}$
     &  1  &  1B    \\
$E1^{(0)}_2=$ & $\frac{1}{N_c}  \left( \xi^{[1,0]} \left( s\ S_c \right)^{[0,0]}\right)^{[1,0]}$
     &  $\frac{1}{N_c}$ & 2B \\
$E1^{(0)}_3=$ & $\frac{1}{N_c}  \left( \xi^{[1,0]} \left( s\ S_c \right)^{[1,0]}\right)^{[1,0]}$
     &  $\frac{1}{N_c}$ &  2B\\
$E1^{(0)}_4=$ & $\frac{1}{N_c}  \left( \xi^{[1,0]} \left( s\ S_c \right)^{[2,0]}\right)^{[1,0]}$
     &  $\frac{1}{N_c}$ &  2B\\
\hline
$E1^{(1)}_1=$ & $\left( \xi^{[1,0]} t\right)^{[1,1]}$
     &  $1$  & \hspace*{.2cm} 1B \hspace*{.2cm}    \\
$E1^{(1)}_2=$ & $\left( \xi^{[1,0]} g\right)^{[1,1]}$
     &  $1$   &   1B\\
$E1^{(1)}_3=$ & $\frac{1}{N_c}  \left( \xi^{[1,0]} \left(  s\ G_c \right)^{[2,1]}\right)^{[1,1]}$
     &  $1$  &  2B  \\
$E1^{(1)}_4=$ & $\frac{1}{N_c}  \left( \xi^{[1,0]} \left(  s\ T_c  \right)^{[1,1]}\right)^{[1,1]}$
     &  $\frac{1}{N_c}$ &  2B\\
$E1^{(1)}_5=$ & $\frac{1}{N_c}  \left( \xi^{[1,0]} \left( s\ G_c  \right)^{[0,1]}\right)^{[1,1]} + \frac{1}{4\sqrt3} \ E1^{(1)}_1$
     &  $\frac{1}{N_c}$ &  2B\\
$E1^{(1)}_6=$ & $\frac{1}{N_c}  \left( \xi^{[1,0]} \left(  s\ G_c  \right)^{[1,1]}\right)^{[1,1]}+ \frac{1}{2\sqrt2} \ E1^{(1)}_2$
     &  $\frac{1}{N_c}$ & 2B\\
 \hline
$M2^{(0)}_1=$ & $\left( \xi^{[1,0]} s\right)^{[2,0]}$
     &  $1$ & 1B   \\
$M2^{(0)}_2=$ & $\frac{1}{N_c}  \left( \xi^{[1,0]} \left( s\ S_c \right)^{[1,0]}\right)^{[2,0]}$
     &  $\frac{1}{N_c}$ & 2B \\
$M2^{(0)}_3=$ & $\frac{1}{N_c}  \left( \xi^{[1,0]} \left( s\ S_c \right)^{[2,0]}\right)^{[2,0]}$
     &  $\frac{1}{N_c}$ & 2B \\
\hline
$M2^{(1)}_1=$ & $\left( \xi^{[1,0]} g\right)^{[2,1]}$
     &  $1$  & 1B    \\
$M2^{(1)}_2=$ & $\frac{1}{N_c}  \left( \xi^{[1,0]} \left(  s\ G_c \right)^{[2,1]}\right)^{[2,1]}$
     &  $1$ & 2B  \\
     $M2^{(1)}_3=$ & $\frac{1}{N_c}  \left( \xi^{[1,0]} \left(  s\ T_c \right)^{[1,1]}\right)^{[2,1]}$
     &  $\frac{1}{N_c}$ &  2B\\
$M2^{(1)}_4=$ & $\frac{1}{N_c}  \left( \xi^{[1,0]} \left(  s\ G_c  \right)^{[1,1]}\right)^{[2,1]}+  \frac{1}{2\sqrt2} \ M2^{(1)}_1$
     &  $\frac{1}{N_c}$ &  2B\\
\hline
$E3^{(0)}_1=$ & $\frac{1}{N_c}  \left( \xi^{[1,0]} \left( s\ S_c \right)^{[2,0]}\right)^{[3,0]}$
     &  $\frac{1}{N_c}$ & 2B  \\
\hline
$E3^{(1)}_1=$ & $\frac{1}{N_c}  \left( \xi^{[1,0]} \left(  s\ G_c  \right)^{[2,1]}\right)^{[3,1]}$
     &  $1$ & 2B \\
\hline \hline
\label{oper}
\end{tabular}
\end{center}
\end{table}
One important check on the basis we have constructed is the counting of the   number of operators
 for each multipole and isospin type.  In general,  we are interested
in transitions of the form $N \gamma \rightarrow B^*$ where $N=p,n$ and $B^*=$ N(1535), N(1520),
 N(1650), N(1700), N(1675), $\Delta(1620)$, $\Delta(1700)$. It is clear that the
following two requirements have to be fulfilled:
i)  isoscalar operators can only contribute to $N \gamma \rightarrow N^*$,
while isovector operators can contribute to both $N \gamma \rightarrow N^*$ and
$N \gamma \rightarrow \Delta^*$, and ii) for each multipole and transition
(independently of the spin/isospin projections) there should be
one and only one  independent element in the operator basis.
Using these, one can proceed to count.
For example,  for $N \gamma \rightarrow N(1535)$ one finds that
there is one independent $E1^{(0)}$ element and one independent $E1^{(1)}$
element.  Similarly, for $N \gamma \rightarrow \Delta(1700)$
we have one independent $E1^{(1)}$ element and one independent $M2^{(1)}$ element. Carrying out this procedure
to the whole $[{\bf 20'},1^-]$ multiplet,
one obtains that the maximum number of independent operators in the different multipoles  are as follows:
$E1^{(0)}$  (4), $E1^{(1)}$ (6), $M2^{(0)}$  (3), $M2^{(1)}$  (4), $E3^{(0)}$  (1), and  $E3^{(1)}$  (1).
Table \ref{oper} shows that  the basis we constructed is consistent  with this count.
The RMEs $\left< J^{*},I^{*}; S^*  || {\cal B}_{n}^{[L,I]} || 1/2,1/2  \right>$
of the  operators in the basis  are shown in Table 2.
They have been obtained using standard angular momentum techniques.
\begin{table}[b]
\setlength{\tabcolsep}{-1pt}
\caption{Reduced matrix elements   of basis operators depicted in Table 1.  The notation $^{2 S^*}N^*_{J^*}$ is used for the nucleon states.  The columns must be multiplied by the corresponding overall factor shown in the last row, where $A\equiv ((1-\frac{1}{N_c})(1+\frac{3}{N_c}))^{1/2}$ and $B\equiv (1-\frac{1}{N_c})^{1/2}$.  The scaling factor $\alpha$ explained in the text is depicted in the last column.}
\vspace*{2mm}
\begin{center}
{\begin{small}
\begin{tabular}{l|ccccccc|c}
\hline \hline
  \hspace*{0.1cm}        &\hspace*{0.0cm}  $^2N^*_{1/2}$  &\hspace*{0.2cm} $^2N^*_{3/2}$  & $^4N^*_{1/2}$  & $^4N^*_{3/2}$ & $^4N^*_{5/2}$ & $\Delta^*_{1/2}$  &$\Delta^*_{3/2}$ & $\alpha$ \\
\hline
$E1^{(0)}_1$ \hspace*{0.1cm}    & $-\sqrt{\frac23}$     & $\sqrt{\frac23}$  &  $-\sqrt{\frac23}$ & $\sqrt{\frac{10}{3}}$   &   $0$  &  $0$    &  $0$ &  $\frac{-3}{\sqrt{10}}$ \\
$E1^{(0)}_2$    & $\frac{-1}{2N_c}$     & $\frac{-1}{N_c}$  &  $0$ & $0$   &   $0$  &  $0$    &  $0$ &  $\sqrt{\frac{3}{2}}$  \\
$E1^{(0)}_3$    & $0$                   & $0$           &  $\frac{-\sqrt{3}}{2N_c}$ & $\frac{\sqrt{15}}{2N_c}$   &   $0$  &  $0$    &  $0$ & $\frac{-2}{\sqrt{5}}$  \\
$E1^{(0)}_4$    & $0$                   & $0$  &  $\frac{\sqrt{5}}{2N_c}$ & $\frac{1}{2N_c}$   &   $0$  &  $0$    &  $0$ & $\frac{-\sqrt{12}}{\sqrt{5}}$  \\
\hline
$E1^{(1)}_1$    & $1$                   & $2$  &  $0$ & $0$   &   $0$  &  $-\sqrt2$    &  $2$ &  $\frac{-\sqrt3}{2\sqrt2}$ \\
$E1^{(1)}_2$    & $\frac{-\sqrt2}{3}$  & $\frac{\sqrt2}{3}$  &  $\frac{1}{3\sqrt2}$ & $\frac{-\sqrt5}{3\sqrt2}$   &   $0$  &  $-\frac{1}{3}$    & $\frac{-1}{3\sqrt2}$ &  $\frac{-3\sqrt{3}}{2\sqrt2}$  \\
$E1^{(1)}_3$    & $0$                   & $0$  &  $\sqrt{\frac{5}{3}}\frac{N_c+2}{4 N_c}$ & $\frac{N_c+2}{4\sqrt3 N_c}$   &   $0$  &  $0$    &  $0$ & $\frac{-36}{5\sqrt{5}}$  \\
$E1^{(1)}_4$    & $\frac{-1}{3\sqrt2 N_c}$ &  $\frac{1}{3\sqrt2 N_c}$   &   $\frac{-2\sqrt2}{3N_c}$  &  $\frac{2\sqrt{10}}{3N_c}$    &  $0$ & $\frac{1}{3N_c}$ &  $\frac{1}{3\sqrt2 N_c}$   &  $\frac{-\sqrt{27}}{\sqrt{40}}$  \\
$E1^{(1)}_5$    &$\frac{1}{4\sqrt3 N_c}$& $\frac{1}{2\sqrt3 N_c}$  &  $0$ & $0$   &   $0$  &  $\frac{1}{\sqrt6 N_c}$    &  $\frac{-1}{\sqrt3 N_c}$ & $\frac{3}{\sqrt{2}}$  \\
$E1^{(1)}_6$    & $0$  & $0$  &  $0$ & $0$   &   $0$  &  $\frac{1}{2\sqrt2 N_c}$    &  $\frac{1}{4 N_c}$ & $-\sqrt{12}$  \\
\hline
$M2^{(0)}_1$    & $0$  & $\sqrt{\frac{10}{3}}$  &  $0$ & $-\sqrt{\frac{2}{3}}$   &   $\frac{3}{\sqrt{2}}$  &  $0$    &  $0$ &  $\frac{-3}{2 \sqrt5}$ \\
$M2^{(0)}_2$    & $0$  & $0$  &  $0$ & $\frac{-\sqrt{3}}{2N_c}$   &   $\frac{9}{4 N_c}$  &  $0$    &  $0$ &  $\frac{-2}{3}$  \\
$M2^{(0)}_3$    & $0$  & $0$  &  $0$ & $\frac{3}{2 N_c}$   &   $\frac{\sqrt{3}}{4N_c}$  &  $0$    &  $0$ & $\frac{-2}{\sqrt{3}}$  \\
\hline
$M2^{(1)}_1$    & $0$  & $\frac{\sqrt{10}}{3}$  &  $0$ & $\frac{1}{3\sqrt{2}}$   &   $\frac{-\sqrt3}{2\sqrt{2}}$  &  $0$    & $\frac{-\sqrt5}{3\sqrt{2}}$  &  $\frac{-\sqrt{27}}{\sqrt{20}}$ \\
$M2^{(1)}_2$    & $0$  & $0$  &  $0$ & $\frac{\sqrt{3}(N_c+2)}{4N_c}$   &   $\frac{N_c+2}{8 N_c}$  &  $0$    &  $0$ &  $\frac{-12}{5}$  \\
$M2^{(1)}_3$    & $0$  & $\frac{\sqrt{5}}{3\sqrt2 N_c}$  &  $0$ & $\frac{-2\sqrt2 }{3 N_c}$   &   $\frac{\sqrt{6}}{N_c}$  &  $0$    &  $\frac{\sqrt{5}}{3\sqrt2 N_c}$ &
$\frac{-\sqrt3}{2\sqrt{2}}$  \\
$M2^{(1)}_4$    & $0$  & $0$  &  $0$ & $0$   &   $0$  &  $0$   & $\frac{\sqrt5}{4 N_c}$  &  $\frac{-2\sqrt{6}}{\sqrt{5}}$ \\
\hline
$E3^{(0)}_1$    & $0$  & $0$  &  $0$ & $0$   &   $\frac{\sqrt{21}}{2\sqrt{2}N_c}$  &  $0$    & $0$  &  $-\sqrt{\frac{6}{7}}$ \\
\hline
$E3^{(1)}_1$    & $0$  & $0$  &  $0$ & $0$   &   $\frac{\sqrt{7}}{4\sqrt{2}}\frac{N_c+2}{N_c}$  &  $0$    & $0$  &  $\frac{-18\sqrt2}{5 \sqrt7}$ \\
\hline \hline
Factor          &$A$& $\frac{-A}{\sqrt{2}}$ &$\frac{-B}{\sqrt{2}}$ &$\frac{-B}{\sqrt{2}}$ &$-\sqrt{\frac 2 3} B$ &$-B$ &$-B$ &
\\
\hline \hline
\end{tabular}
\end{small}}
\end{center}
\end{table}

\begin{table*}[t]
\caption{Helicity amplitudes (in units of $10^{-3} {\rm Gev}^{-1/2}$) for the fits in
Table \ref{coe}.   The sign $\eta$ is indicated in the last column. In the fits we have set $\xi=-1$, and $\kappa=+1$.  Numbers in parenthesis indicate the individual contribution to the
total $\chi^2$. }
\vspace*{2mm}
\label{heli}
\begin{center}
\begin{tabular}{ccrrrrr}
\hline \hline
Amplitude \hspace{.5cm}  & \hspace{.5cm} Empirical \hspace{.5cm} &\hspace{.6cm} LO  & \hspace{.3cm}NLO1
 & \hspace{.8cm}NLO2 &\hspace{.8cm} NLO3 & \hspace{.8cm}$\eta$\hspace{.8cm}\\  \hline
$A^p_{1/2}[N(1535)]$  & $+90\pm 30$
&$76(0.2)$   & $ 90$  &  $111(0.5)$  & $86(0.0)$
& $-\xi$\\
$A^n_{1/2}[N(1535)]$  & $-46\pm 27$
& $-54(0.1)$ & $-46$  &  $-78(1.4)$  & $-72(0.9)$
& $-\xi$\\
\hline
$A^p_{1/2}[N(1520)]$  & $-24\pm 9$
& $-25(0.0)$ & $-24$  &  $-20(0.2)$  & $-16(0.8)$
& $-1$\\
$A^n_{1/2}[N(1520)]$  & $-59\pm 9$
& $-6(8.8)$  & $-59$  &  $-46(1.9)$  & $-43(3.1)$
& $-1$\\
$A^p_{3/2}[N(1520)]$  & $+166\pm 5$
& $66(4.0)$ &  166  & 163(0.4)  &  162(0.7)
& $-1$\\
$A^n_{3/2}[N(1520)]$  & $-139\pm 11$
& $-55(4.0)$    &  $-139$   &  $-143(0.1)$  &  $-135(0.1)$
& $-1$\\
\hline
$A^p_{1/2}[N(1650)]$  & $+53\pm 16$
&  45(0.3) &     53   &   52(0.0) &    39(0.8)
& $\xi$ \\

$A^n_{1/2}[N(1650)]$  & $-15\pm 21$
& $-12(0.0)$ &   $-15$ &  $-20(0.1)$  &  $-25(0.2)$
& $\xi$\\
\hline
$A^p_{1/2}[N(1700)]$  & $-18\pm 13$
&$-18(0.0)$  &   $-18$ &  $-20(0.0)$  &  $26(11.7)$
& $\kappa$\\
$A^n_{1/2}[N(1700)]$  & $0\pm 50$
&$41(0.7)$   &      0 &   $47(0.9)$   &  $-13(0.1)$
& $\kappa$\\
$A^p_{3/2}[N(1700)]$  & $-2\pm 24$
&$1(0.0)$     &  $-2$  &   $-10(0.1)$  &  $-46(3.3)$
& $\kappa$\\
$A^n_{3/2}[N(1700)]$  & $-3\pm 44$
&47(1.3)     &  $-3$  &   47(1.3)     &   61(2.1)
& $\kappa$\\
\hline
$A^p_{1/2}[N(1675)]$  & $+19\pm 8$
&15 (0.3)    &   19   &   8(2.0)      &   2(4.4)
& $-1$\\
$A^n_{1/2}[N(1675)]$  & $-43\pm 12$
&$-45(0.0)$  &  $-43$ &  $-50(0.4)$   &  $-43(0.0)$
& $-1$\\
$A^p_{3/2}[N(1675)]$  & $+15\pm 9$
&$10(0.3)$   &  15    &  $11(0.2)$    &  $3(1.7)$
& $-1$\\
$A^n_{3/2}[N(1675)]$  & $-58\pm 13$
&$-53(0.1)$     &  $-58$    &$-71(1.0)$  &   $-61(0.0)$
& $-1$\\
\hline
$A^N_{1/2}[\Delta(1620)]$ & $+27\pm 11$
& $53(5.7)$  &     27  &      32(0.2)   &    81(24.5)
& $-\xi$\\
\hline
$A^N_{1/2}[\Delta(1700)]$ & $+104\pm 15$
& 80(0.6)  &     104 &     108(0.1)   &    90(0.9)
& $+1$\\
$A^N_{3/2}[\Delta(1700)]$ & $+85\pm 22$
&70(0.3)   &     85  &     112(1.5)   &    67(0.6)
& $+1$\\
\hline \hline\end{tabular}
\end{center}
\end{table*}

One more important input needed from the strong
transitions is the  sign $\eta(B^*)$ that appears in Eqs. (3)--(4).  That sign
is obtained from the strong amplitude  for $ B^*  \rightarrow  \pi N$, and is
given in terms of the corresponding RME defined in Ref. \cite{GSS1} by
\beq
\eta(B^*) =(-1)^{J^*-\fr 12}\;
{\rm sign}(\langle \ell_\pi \; N\parallel H_{\rm QCD}\parallel J^*\;I^*\rangle),
\eeq
where $\ell_\pi$ corresponds to the pion partial wave.
Note that the sign $\eta$ can be  determined   up to an overall sign for
each pion partial wave, which cannot be fixed by strong transitions alone.  Since the partial waves involved in our case are  $S$ and $D$
waves, we have one extra relative sign, which we will call $\xi$
as customary \cite{GK74,Babcock:1975bw}. In addition, the analysis of the
strong transitions
gives two consistent but different results for the mixing angle $\theta_3$. The values
(in radians)
$\theta_3=2.82$ and $\theta_3=2.38$ cannot be distinguished from the  strong fits. One finds that
some of the
$\eta$ signs are different for these two values. We take into account this with an extra
sign factor $\kappa$,
which is equal to $+1$$~(-1)$  for $\theta_3=2.82$$~(2.38)$.

Table \ref{heli} displays  the empirically known helicity amplitudes taken from
\cite{PDG06} along with the   strong sign $\eta$, and  the amplitudes resulting
from the fits to be discussed in the next section.

\section{Analysis and Results }

In this section we present and analyze the different fits to the helicity amplitudes.
The coefficients to be fitted $g^{[L, \, I]}_{n,L'}(\omega)$ are expressed by  including the
barrier penetration factor:  $g^{[L, \, I]}_{n,L'} \times (\omega/\Lambda)^{L'}$,
where $L'=0$ for $E1$ operators and $L'=2$ for $M2$ and $E3$ operators.
Throughout we will choose the scale $\Lambda=m_\rho$.
We performed several LO and NLO  fits.
A first analysis concerns the choices left
by the values of the mixing angle $\theta_3$, and the signs $\xi$ and $\kappa$. Using
all the  LO  operators, the choices are made by considering  the $\chi^2$ for all possibilities.
The sign $\xi=-1$ is strongly favored.    This is
in agreement with an old  determination  based on the single-quark-transition model \cite{GK74,Babcock:1975bw}.
The second choice that is favored, although less markedly than the one for $\xi$, is $\theta_3=2.82$. Finally, for
$\kappa$ there is no indication of a preference from the fits; for the sake a definiteness we will take $\kappa=+1$ in our fits.
This latter  sign basically depends on strong
amplitudes which are small and have large relative errors,  which
imply  that its determination is subject to a degree of
uncertainty. The helicity amplitudes show here their importance by allowing to determine the relative sign $\xi$ between
the strong $S$ and $D$ wave amplitudes, and by selecting between the two possible values of  $\theta_3$ consistent with the
strong transitions. Note that $\theta_3=2.82$ corresponds to \lq\lq small\rq\rq mixing, while 2.32 corresponds to \lq\lq large\rq\rq mixing.
 A simultaneous fit of strong
transitions and photoproduction amplitudes is the best  way of
extracting the mixing angles. This will be carried out in a future
project \cite{GMS2}.

As already mentioned, the helicity amplitudes resulting from the fits we have carried
out are given in Table \ref{heli};  the corresponding fit coefficients are displayed
in Table \ref{coe}.  In the fits we expand the operator matrix elements in powers of $1/N_c$
to the order corresponding to the fit. In the LO fits,  we have
set the errors in the input helicity amplitudes to be equal to 0.3
of the value of the helicity amplitude or the experimental value
if this is larger. The point of  this is to test whether or
not the LO analysis is consistent in the sense that it gives a $\chi^2$ per degree of freedom
($\chi^2_{\mathrm{\rm dof}}$)  close to unity. For the NLO fits,  we of course use
the empirical errors.

\begin{table}
\caption{Results for the dimensionless coefficients  $g^{[L,\, I]}_{n,L'}$  from different fits.
Two partial NLO fits are given.  Fit NLO2 keeps the minimum number of dominant operators needed for
$\chi^2_{\mathrm{\rm dof}}\leq 1$, and fit NLO3 only keeps 1-body operators.  }
\vspace*{2mm}
\label{coe}
\begin{center}
\begin{tabular}{ccccc}
\hline\hline
\hspace*{-.08cm}  Operator \hspace*{.008cm}   & \hspace*{.008cm} LO   \hspace*{.08cm}           &
\hspace*{.08cm}  NLO1     \hspace*{.08cm}   & \hspace*{.08cm} NLO2  \hspace*{.08cm}           &
\hspace*{.08cm}  NLO3     \hspace*{.08cm} \\
\hline
$E1^{(0)}_1$    & $-0.36 \pm 0.19$  &    $-0.34 \pm   0.22$  &  $-0.34 \pm 0.15$ & $-0.15\pm 0.14$ \\
$E1^{(0)}_2$    &                   &    $ 0.52 \pm   0.62$  &                   &                 \\
$E1^{(0)}_3$    &                   &    $ 1.02 \pm   0.85$  &                   &                 \\
$E1^{(0)}_4$    &                   &    $ 0.50 \pm   0.63$  &                   &                 \\
$E1^{(1)}_1$    & $2.34  \pm  0.31$ &    $ 3.03 \pm   0.20$  &  $ 3.54 \pm 0.13$ & $ 3.26\pm 0.22$ \\
$E1^{(1)}_2$    & $-0.68 \pm  0.36$ &    $ 0.40 \pm   0.27$  &                   & $ 0.21\pm 0.25$ \\
$E1^{(1)}_{3}$& $ 0.41 \pm  0.53$ &    $-0.21 \pm   0.41$  &                   &                 \\
$E1^{(1)}_{4}$&                   &    $-1.95 \pm   1.42$  &                   &                 \\
$E1^{(1)}_{5}$&                   &    $-0.18 \pm   0.90$  &                   &                 \\
$E1^{(1)}_{6}$&                   &    $ 4.17 \pm   0.89$  &  $ 3.92 \pm 0.77$ &                 \\
$M2^{(0)}_1$    & $ 0.76 \pm  0.21$ &    $ 1.52 \pm   0.32$  &  $ 1.27 \pm 0.17$ & $ 1.21\pm 0.17$ \\
$M2^{(0)}_2$    &                   &    $-1.22 \pm   1.34$  &                   &                 \\
$M2^{(0)}_3$    &                   &    $-1.18 \pm   1.75$  &                   &                 \\
$M2^{(1)}_1$    & $ 3.02 \pm  0.62$ &    $ 3.81 \pm   0.56$  &  $ 3.95 \pm 0.40$ & $ 4.69\pm 0.37$ \\
$M2^{(1)}_{2}$& $-3.11 \pm  1.00$ &    $-2.33 \pm   1.12$  &  $-2.73 \pm 0.62$ &                 \\
$M2^{(1)}_{3}$&                   &    $-0.15 \pm   1.13$  &                   &                 \\
$M2^{(1)}_{4}$&                   &    $-1.49 \pm   2.38$  &                   &                 \\
$E3^{(0)}_1$    &                   &    $ 0.34 \pm   0.83$  &                   &                 \\
$E3^{(1)}_1$    & $ 0.75 \pm  0.89$ &    $ 0.35 \pm   0.53$  &                   &                 \\
\hline\hline
$\rm dof$           		&$11$               & $0$           &13                 &$14$           \\
\hline
$\chi_{\rm dof}^2$  &$2.42$             & $-$           &0.94               &$4.00$         \\
\hline\hline
\end{tabular}
\end{center}
\label{}
\end{table}

We now proceed to discuss the results.
\begin{itemize}
\item The LO  fit  shows a $\chi^2_{\mathrm{\rm dof}}$ of 2.42. This indicates that there are NLO
effects to be taken into account for a satisfactory fit.  The main deficiencies are in fitting of the $N(1520)$ and
the $\Delta(1620)$ amplitudes  as one can readily ascertain from their individual contributions to the total $\chi^2$
(numbers in parenthesis in Table \ref{heli}).
If one keeps only the LO operators with the largest coefficients (say coefficients
bigger than 2), the $\chi^2_{\mathrm{\rm dof}}$  does not change much from the one obtained with all LO operators.
Notice that one 2-body LO operator seems to be significant, namely $M2_2^{(1)}$. We have checked that a fit taking
$\kappa=-1$ leads to similar results except that the coefficient of $M2_2^{(1)}$ results to be only 40\% of the case
$\kappa=+1$. If indeed 2-body operators should give small effects, then this would be a way to discriminate about the
sign $\kappa$. In fact, a LO fit using only 1-body operators gives respectively $\chi^2_{\mathrm{\rm dof}}=2.48$  and
2.12 for $\kappa=+1$ and $-1$.

\item One can perform a LO fit motivated by  the  single-quark-transition
model  \cite{GK74,Babcock:1975bw},  which is also commonly used in quark model calculations. In that model,
the photon only couples to the excited quark with a fixed ratio for
the isoscalar versus the isovector coupling as given by the bare quark charges.
Here this is achieved  by locking 1-body operators as follows:
$(\frac{1}{6}E1_1^{(0)}+\;E1_2^{(1)})$, $(\frac{1}{6}M2_1^{(0)}+M2_1^{(1)})$, and
$E1_2^{(1)}$ whose isoscalar counterpart does not appear in the operator basis because it  is spin-flavor singlet.
The fit  has  $\chi^2_{\rm dof}\sim 2.5$ at LO, which is similar to the result with unlocked operators, thus indicating that at LO  one cannot draw a clear conclusion.

\item
As it is well known, in the single-quark-transition model the so-called Moorhouse selection  rule \cite{Clo79} holds. That rule states that the amplitudes for photoexcitation of protons to $^4N^*$ states   vanish.
In the present analysis,  the
rule is violated by  the unlocking of the 1-body operators, and by 2-body operators. At the level of physical states, the rule
tends to suppress the amplitudes $p\gamma \to N(1650), \;N(1700),\; {\rm and}\;N(1675)$.  In the first
two cases, the mixing angles $\theta_1$ and $\theta_3$  work against that suppression as they give to
these states a component $^2N^*$. In the case of  $N(1675)$,  the rule turns out to be mostly violated by 2-body
effects, at least for $\kappa=+1$.

\item The NLO  order fit NLO1,  involves all operators in the basis. It   gives values for
the  coefficients of the LO operators  which are, within the expected deviations  from
$1/N_c$ counting, consistent with the values  obtained in the LO fits.
  Moreover, none of the coefficients of the NLO
operators has a magnitude larger than that of the largest  LO coefficients.  This  is a strong indication of the
consistency of the $1/N_c$ expansion.   We find that this
consistency is more clearly manifested here than in the case of
the positive parity baryons  analyzed in \cite{GSelm}.
 From the magnitude  of the  coefficients,  it is obvious that only a few NLO operators are needed
for a consistent fit.  In fact, as shown by the fit NLO2 in Table
\ref{coe},  a consistent fit is obtained with only five LO  and
one NLO  operators.  Of these dominant
operators four are one-body and LO, and two are
two-body with one of them LO and the other  NLO.
Note also that none of the 2-body $E3$ operators is required.
It is remarkable  that out of eleven NLO
operators only one is  essential for obtaining consistent fits.  At this point it is
important to mention that  many of  the empirical
amplitudes have errors  that are larger than  what is needed  for an
accurate  NLO  analysis.  It is for this reason that one cannot draw a more precise
NLO picture which could unveil  the role of other   operators.
\item To test  for deviations from the  single-quark-transition model at NLO,  we have performed a NLO fit including all operators with locked   the 1-body operators.   The result is a $\chi^2_{\rm dof}\sim 2.5$, which gives a good indication  that there are deviations
from that model.
\item The  fit  NLO3  depicted in Table 4  including  only 1-body operators  gives  rather large  $\chi^2_{\rm dof}$, with a similar
result  for a 1-body fit with locked operators.  One can conclude that, although the gross features of the set of helicity amplitudes
are described by 1-body operators, the deviations  can   be pinpointed quite clearly, in particular the need for 2-body effects.

\item
The dominant operator in terms of the magnitude of its contributions is  $E1_1^{(1)}$,
as it can be seen from Table  5,  which depicts the partial contribution to each amplitude by 
the operators included in the fit NLO2.
This operator is expected to dominate in a non-relativistic quark model as it corresponds to the
usual orbital electric dipole transition.  The contributions of the other relevant $E1$ and $M2$ operators, 
needed for a consistent fit,  turn  out to be rather similar in magnitude.

\end{itemize}

~

It is instructive to briefly discuss the individual helicity amplitudes, as they differ very significantly
in the type of contributions involved. For this discussion,  we take fit NLO2, which contains the  most
significant contributions.  As already mentioned,  the individual contributions by the various operators
to the helicity amplitudes are shown in Table 5.

\begin{itemize}
\item $N(1535)$: The amplitudes are not very well established, with various analyses giving significantly different
results \cite{PDG06,Arndt}. One can however establish that $E1_1^{(1)}$ plays an important role,  in particular because its coefficient   is primarily  determined by other better known amplitudes. For this reason, we find that it is very difficult to
reconcile the values obtained for the amplitudes on $p$ and $n$  which result  from the analysis carried out  in Ref. \cite{Arndt}.

\item  $N(1520)$: This as, well as $N(1700)$, receive several contributions $E1$ and $M2$, which involve some
important cancellations.  One manifestation of such cancellations is in the $\lambda=1/2$ amplitudes in which
the isoscalar component turns out to be larger than the isovector one (only case where this occurs). In the
quark model such a cancellation seems unproblematic to be explained \cite{Capstick:2000qj}, and thus it can be understood
in simple terms.  On the other hand, the $\lambda=3/2$ amplitudes are dominated by the operator $E1_1^{(1)}$, with
small contributions from other operators, and a particularly small total isosinglet component.

\item $N(1650)$: The $p\gamma$ amplitude  would vanish in the limit in  which the Moorhouse rule is valid.
The dominant effect driving this  amplitude  is the mixing by the angle  $\theta_1$. One can check that the effect of
unlocking operators gives small contributions, and in particular tends  to reduce the Moorhouse allowed
$n\gamma$ amplitude  (for the latter there are however some discrepancies between different analyses \cite{PDG06,Arndt}).

\item $N(1700)$: These amplitudes are poorly known empirically, as they seem to be small (some of them on the grounds
of the Moorhouse rule). In addition several operators contribute, which according to our analysis will tend to have
large cancellations. Thus, one expects that a clear understanding of the physics contained in this case will not be easy.

\item $N(1675)$:  These are the only amplitudes admitting $E3$ contributions,  and show through the fit that they are irrelevant.
Note that the $E3$ operators are 2-body.  In this case the $p\gamma$ amplitudes  only proceed because of  violations to the
Moorhouse rule due to  the unlocking of 1-body operators and  due to 2-body operators.  We find that  the main  contribution  is due to the
2-body LO operator $M2_2^{(1)}$.  On the other hand the unsuppressed $n \gamma$ amplitudes are dominated by the $M2$ 1-body contributions.

\item $\Delta(1620)$: Various analyses are inconsistent with each other, but all of them strongly indicate that this helicity amplitude
is small. It is an interesting amplitude, because it receives a large $E1$ contribution from $E1_1^{(1)}$, and the only way to have a small
amplitude is  to have a large cancellation.  In our analysis that cancellation is shown to come from the 2-body operator $E1_6^{(1)}$; in fact
the need for this cancellation largely determines in the fit the importance of that operator.  Taken at face value, this is a strong indication
for 2-body effects.
In the single-quark-transition model, as well as in quark models \cite{Capstick:2000qj} one finds that the calculated
amplitude is much larger than the empirical one.
This is  due to the absence  of  the 2-body effects in those models.

\item $\Delta(1700)$: These are among the most clearly established and understood amplitudes. $E1_1^{(1)}$ plays
the dominant role, with the other two operators $M2_1^{(1)}$ and  $E1_6^{(1)}$ giving contributions of similar magnitude.
Since the 1-body  LO operators already give a good description, it is not surprising that these amplitudes are well described
in the quark model \cite{Capstick:2000qj}.
\end{itemize}

At this point we can compare our analysis with that of Carlson and Carone \cite{carlson}. We have checked that their set of operators,
eleven in total, corresponds  to a subset of our operator basis,  which can be obtained by locking several pairs of operators using the
isoscalar to isovector ratio of the electric charge operator as we explained earlier. In this case,  1- as well as 2-body operators are locked.  A  fit with that set of locked operators gives a $\chi^2_{\rm dof} \sim 3.2$.
This result clearly   indicates the necessity for  the more general basis
we use in this work.  However, one should  emphasize that the main features of most helicity amplitudes are  obtained in the analysis
of  Ref.  \cite{carlson}.  Another point where we differ with Ref.   \cite{carlson}  is in the mixing angles:  in our analysis we take the mixing angles  from the strong
decays,  while in Ref.  \cite{carlson} some of the fits include fitting the mixing angles. Their mixing angles are somewhat different from ours,  leaving an open issue which should be sorted out.    We plan to carry out simultaneous fits of strong decays and helicity amplitudes
\cite{GMS2},  from where we expect to extract more reliable values for the mixing angles.

\begin{table}[h]
\caption{Partial contributions to the helicity amplitudes by the different operators.
 This table corresponds to the NLO2 fit. }
 \vspace*{2mm}
\label{par}
\begin{center}
\begin{tabular}{c|ccccccc}
\hline\hline
     Amplitude   &\hspace*{-0.026cm} $E1^{(0)}_1$ \hspace*{-0.0512cm}
                          &\hspace*{-0.026cm} $E1^{(1)}_1$ \hspace*{-0.0512cm}
                          &\hspace*{-.026cm} $E1^{(1)}_6$ \hspace*{-0.0512cm}
                          &\hspace*{-0.026cm} $M2^{(0)}_1$ \hspace*{-0.0512cm}
                          &\hspace*{-0.026cm} $M2^{(1)}_1$ \hspace*{-0.0512cm}
                          &\hspace*{-0.026cm} $M2^{(1)}_2$ \hspace*{-0.0512cm}
                          &\hspace*{-0.024cm} Total \hspace*{-0.002cm}
\\ \hline  \hline
$A^p_{1/2}[N(1535)]$      &$  17 $    &$  95 $      &$   0 $    &$   0 $       &$   0 $    &$   0 $   &$ 111 $  \\
$A^n_{1/2}[N(1535)]$      &$  17 $    &$ -95 $      &$   0 $    &$   0 $       &$   0 $    &$   0 $   &$ -78 $  \\
\hline
$A^p_{1/2}[N(1520)]$      &$  -4 $    &$  70 $      &$   0 $    &$ -29 $       &$ -46 $    &$ -11 $   &$ -20 $  \\
$A^n_{1/2}[N(1520)]$      &$  -4 $    &$ -70 $      &$   0 $    &$ -29 $       &$  46 $    &$  11 $   &$ -46 $  \\
$A^p_{3/2}[N(1520)]$      &$  -7 $    &$ 120 $      &$   0 $    &$  17 $       &$  26 $    &$   6 $   &$ 163 $  \\
$A^n_{3/2}[N(1520)]$      &$  -7 $    &$-120 $      &$   0 $    &$  17 $       &$ -26 $    &$  -6 $   &$-143 $  \\
\hline
$A^p_{1/2}[N(1650)]$      &$  16 $    &$  36 $      &$   0 $    &$   0 $       &$   0 $    &$   0 $   &$  52 $  \\
$A^n_{1/2}[N(1650)]$      &$  16 $    &$ -36 $      &$   0 $    &$   0 $       &$   0 $    &$   0 $   &$ -20 $  \\
\hline
$A^p_{1/2}[N(1700)]$      &$  11 $    &$ -21 $      &$   0 $    &$   2 $       &$  32 $    &$ -44 $   &$ -20 $  \\
$A^n_{1/2}[N(1700)]$      &$  11 $    &$  21 $      &$   0 $    &$   2 $       &$ -32 $    &$  44 $   &$  47 $  \\
$A^p_{3/2}[N(1700)]$      &$  20 $    &$ -36 $      &$   0 $    &$  -1 $       &$ -18 $    &$  26 $   &$ -10 $  \\
$A^n_{3/2}[N(1700)]$      &$  20 $    &$  36 $      &$   0 $    &$  -1 $       &$  18 $    &$ -26 $   &$  47 $  \\
\hline
$A^p_{1/2}[N(1675)]$      &$   0 $    &$   0 $      &$   0 $    &$ -21 $       &$  19 $    &$  10 $   &$   8 $  \\
$A^n_{1/2}[N(1675)]$      &$   0 $    &$   0 $      &$   0 $    &$ -21 $       &$ -19 $    &$ -10 $   &$ -50 $  \\
$A^p_{3/2}[N(1675)]$      &$   0 $    &$   0 $      &$   0 $    &$ -30 $       &$  27 $    &$  14 $   &$  11 $  \\
$A^n_{3/2}[N(1675)]$      &$   0 $    &$   0 $      &$   0 $    &$ -30 $       &$ -27 $    &$ -14 $   &$ -71 $  \\
\hline
$A^N_{1/2}[\Delta(1620)]$ &$   0 $    &$  85 $      &$ -53 $    &$   0 $       &$   0 $    &$   0 $   &$  32 $  \\
\hline
$A^N_{1/2}[\Delta(1700)]$ &$   0 $    &$  57 $      &$  18 $    &$   0 $       &$  32 $    &$   0 $   &$ 108 $  \\
$A^N_{3/2}[\Delta(1700)]$ &$   0 $    &$  99 $      &$  31 $    &$   0 $       &$ -19 $    &$   0 $   &$ 112 $ \\
\hline \hline
\end{tabular}
\end{center}
%\label{}
\end{table}

\section{Summary}

The aim of this work was to  extend the $1/N_c$ expansion analysis of baryon photoproduction helicity amplitudes to the negative
parity baryons,  improving  on the  approach used in earlier work \cite{carlson}. The most important outcome of the analysis is that
the expected hierarchies implied by the $1/N_c$ power counting are respected.  Another important aspect is that only a reduced number
of the operators in the basis  turn out to be  relevant.   Several of those operators can be easily identified with those in quark models,
but there are also 2-body operators not included in quark models which are necessary for an accurate description of the empirical helicity
amplitudes.  With this analysis one can select between the two possible values of the mixing angle $\theta_3$ which are consistent with
strong decays, as well as the relative sign $\xi$  between the $S$ and $D$-wave strong amplitudes.
A comprehensive analysis that includes strong and helicity amplitudes will  further refine the results of this work, and will be presented
elsewhere.
\vspace{0.6cm}

This work was supported by DOE (USA) through contract DE-AC05-84ER40150,
and by   NSF (USA) through grants PHY-0300185 and  PHY-0555559 (JLG),  by CONICET (Argentina)
grant  PIP 6084,  and by ANPCyT  (Argentina) grant  PICT04 03-25374 (NNS),  and by the Institut
Interuniversitaire des Sciences Nucl\'eaires (Belgium) (NM).

\end{document}